# Choosing Between Methods of Combining $p$-values

Nicholas A. Heard

Department of Mathematics, Imperial College London

and

Patrick Rubin-Delanchy

School of Mathematics, University of Bristol

## Abstract

Combining $p$-values from independent statistical tests is a popular approach to meta-analysis, particularly when the data underlying the tests are either no longer available or are difficult to combine. A diverse range of $p$-value combination methods appear in the literature, each with different statistical properties. Yet all too often the final choice used in a meta-analysis can appear arbitrary, as if all effort has been expended building the models that gave rise to the $p$-values. Birnbaum (1954) showed that any reasonable $p$-value combiner must be optimal against some alternative hypothesis. Starting from this perspective and recasting each method of combining $p$-values as a likelihood ratio test, we present theoretical results for some of the standard combiners which provide guidance about how a powerful combiner might be chosen in practice.

# 1 Introduction

Suppose $p_1, \ldots, p_n$ are $p$-values obtained from $n$ independent hypothesis tests. If the underlying test statistics $t_1, \ldots, t_n$ have absolutely continuous probability distributions under their corresponding null hypotheses, the joint null hypothesis for the $p$-values is

$$H_0 : p_i \sim \mathrm{U}[0,1], \quad i = 1, \ldots, n. \tag{1}$$

The broadest possible alternative hypothesis is

$$H_1 : p_i \sim f_{1,i}, \quad i = 1, \ldots, n, \tag{2}$$

where $f_{1,i}$ is a possibly unknown, non-increasing density with support on the unit interval. To see that $f_{1,i}$ can be assumed to be non-increasing without loss of generality, see Birnbaum (1954). Under this broad alternative, he showed that any test statistic for combining the $p$-values which is monotonic in the $p$-values is admissible, in the sense that there exists a combination of densities $f_{1,i}$ for which this combiner is optimal.

The six most fundamental or commonly used statistics for combining $p$-values are attributed as follows: $S_\mathrm{F} = \sum_{i=1}^n \log p_i$ (Fisher, 1934), $S_\mathrm{P} = -\sum_{i=1}^n \log(1-p_i)$ (Pearson, 1933), $S_\mathrm{G} = S_\mathrm{F} + S_\mathrm{P} = \sum_{i=1}^n \log\{p_i/(1-p_i)\}$ (Mudholkar and George, 1979), $S_\mathrm{E} = \sum_{i=1}^n p_i$ (Edgington, 1972), $S_\mathrm{S} = \sum_{i=1}^n \Phi^{-1}(p_i)$ (Stouffer et al., 1949), where $\Phi$ is the standard normal cumulative distribution function, and $S_\mathrm{T} = \min(p_1, \ldots, p_n)$ (Tippett, 1931). Clearly, each is monotonic in the $p$-values, and therefore optimal in some setting. However, little attention has been afforded to highlighting precisely the settings in which these combiners are optimal, and thereby determining their suitability for an application at hand.

Instead, the widespread adoption of the above statistics for combining $p$-values can be largely attributed to their simplicity and mathematical convenience: Under $H_0$, $-2S_\mathrm{P}$ and $-2S_\mathrm{F}$ are both distributed as $\chi^2_{2n}$; $S_\mathrm{G}$ and $S_\mathrm{E}$ have slightly more awkward closed-form distributions which can be well approximated by Gaussian distributions for large $n$; $S_\mathrm{S}$ is $\mathrm{N}(0, n)$; and $S_\mathrm{T} \sim \mathrm{Beta}(1, n)$. A meta-analysis combined $p$-value is therefore trivial to obtain for each of these statistics, but they can differ substantially. Figure 1 demonstrates how each method combines two $p$-values $(p_1, p_2)$ from $[0,1]^2$ into a single significance level. Tippett's and Fisher's methods are clearly more sensitive to the smallest $p$-value, while Pearson's method is most sensitive to the largest $p$-value. George's, Edgington's and Stouffer's methods can be seen as compromises; for example, in all three cases $(0, 1) \mapsto 0.5$. There are also some similarities: there is an approximate equivalence between Pearson's and Edgington's methods for combining small $p$-values, due to the Taylor series approximation $-\log(1 - x) \approx x$ for small $x > 0$. Additionally, the well-understood similarity between the logistic and Gaussian distributions causes George's and Stouffer's methods to be very similar, except at the extremes where the heavier tail of the logistic distribution translates to lower sensitivity to extreme $p$-values.

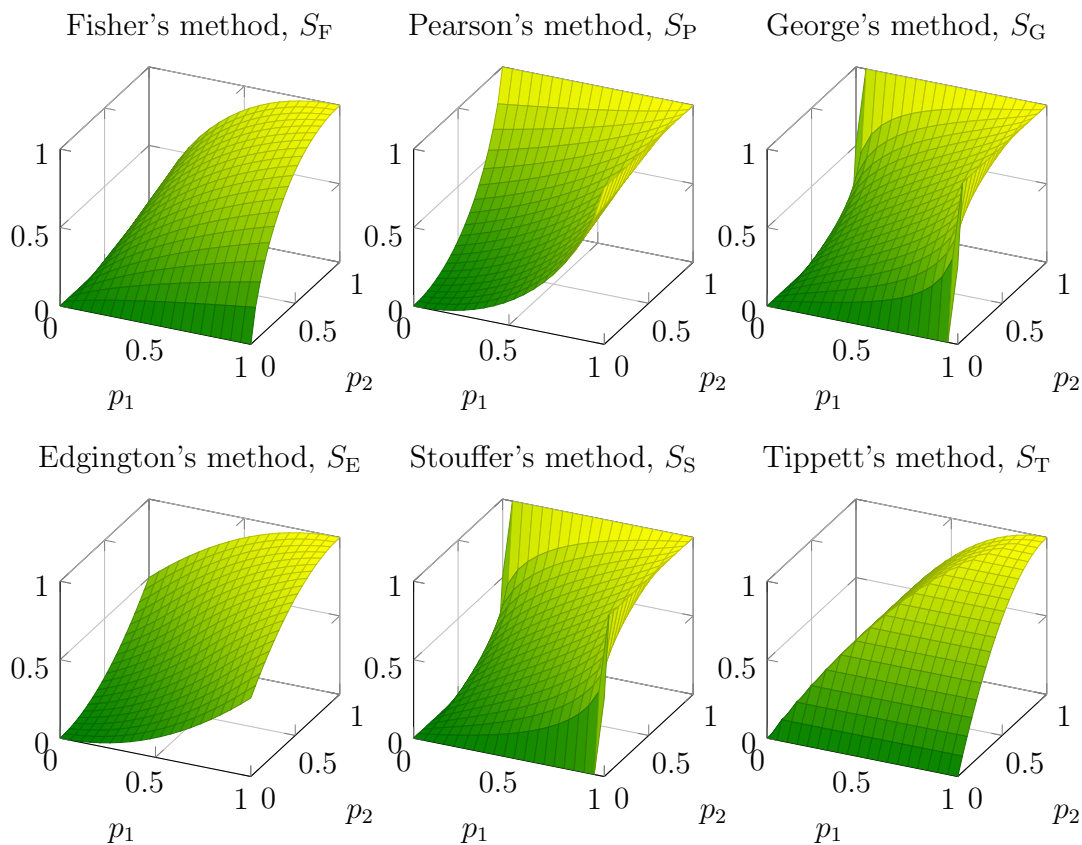

Figure 1: Significance levels from different methods of combining two $p$-values.

Which combination method should be used? Several simulation studies investigating this question have been published; see, for example, Loughin (2004) and Kocak (2017). However, each of the combiners discussed in those studies and here can be interpreted as a likelihood ratio test for a range of different, more specific variations of $H_1$ than (2). By the Neyman–Pearson lemma, it is therefore possible to identify settings in which each method of combining $p$-values provides the uniformly most powerful test.

## 2 Alternative Hypotheses and Likelihood Ratio Tests

### 2.1 The Beta Distribution

In empirical simulation studies of combining $p$-values, the Beta$(a,b)$ density

$$f_1(p) \propto \mathbb{I}_{[0,1]}(p) \; p^{a-1}(1-p)^{b-1}, \quad a,b > 0,$$

has provided a natural choice for specifying an alternative density for the $p$-values, since it has the correct support and is the conjugate prior for event probabilities in Bayesian analysis. As noted in Section 1, $f_1(p)$ can be assumed to be non-increasing in $p$, which for the beta distribution corresponds to requiring $a \in (0,1]$ and $b \in [1,\infty)$.

**Proposition 1.** *Consider an alternative hypothesis for p-values*

$$H_1 : p_i \sim \text{Beta}(a,b), \quad i = 1, \ldots, n, \tag{3}$$

*with $a \in (0,1]$, $b \in [1,\infty)$ and $a < b$. Let $w = (1-a)/(b-a)$. Then the uniformly most powerful test statistic for combining $p_1, \ldots, p_n$ is of the form*

$$w \sum_{i=1}^{n} \log p_i - (1-w) \sum_{i=1}^{n} \log(1 - p_i). \tag{4}$$

*Proof.* Since the $p$-values have unit density under $H_0$, the log-likelihood ratio is $-\sum_{i=1}^{n} \log f_1(p_i) = (1-a) \sum_{i=1}^{n} \log p_i - (b-1) \sum_{i=1}^{n} \log(1-p_i)$. The rescaling of the weights to $w, 1-w$ with $0 \leq w \leq 1$ highlights that the same test statistic is optimal for an infinite collection of beta distributions with the same ratio $(1-a)/(b-a)$. □

Therefore under the alternative hypothesis (3), the optimal $p$-value combiner is a weighted combination of $S_\text{F}$ and $S_\text{P}$. In particular, $S_\text{F}$ is optimal for an alternative where $p$-values are Beta$(a,1)$, $S_\text{P}$ is optimal against Beta$(1,b)$, and $S_\text{G}$ is optimal for Beta$(a,b)$ whenever $a + b = 2$. The left panel of Figure 2 shows example densities for each. The Beta(0.5,1) density, suited to $S_\text{F}$, has the steepest acceleration to infinity as $p \to 0$, consistent with the remark in Section 1 that Fisher's method is sensitive to very small $p$-values.

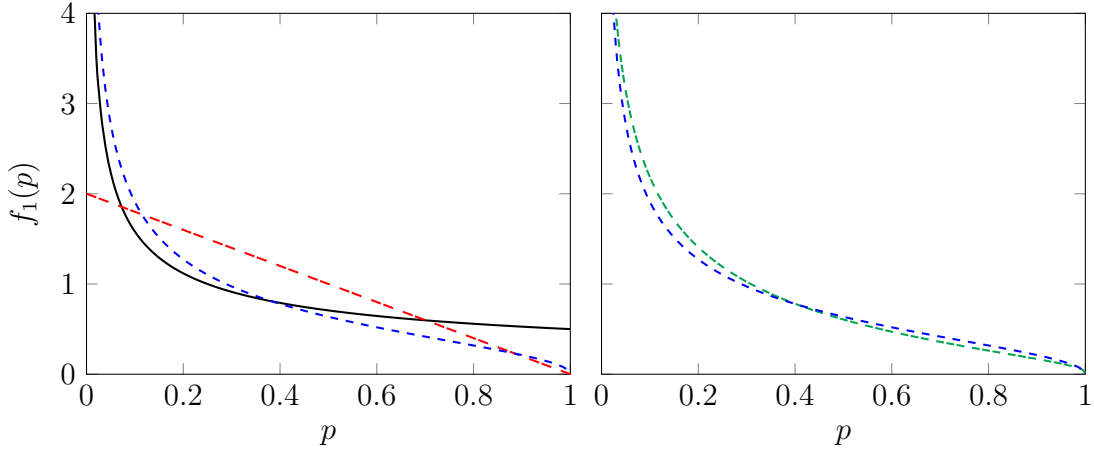

Figure 2: Example beta distribution densities. Left: Cases for which $p$-values are optimally combined using Fisher's method (Beta$(0.5, 1)$, —), Pearson's method (Beta$(1, 2)$, - - -) or George's method (Beta$(0.5, 1.5)$, - - -). Right: Comparison of Beta$(0.5, 1.5)$ density with the density $f_1(p) = \exp\{-\Phi^{-1}(p) - 0.5\}$ (—·—) of $p$-values from Example 2.3 for $\mu = -1$.

**Example 2.1.** *Suppose $t_1, \ldots, t_n$ are independent $\text{Exp}(\lambda)$ inter-arrival times from an homogeneous Poisson process with a null hypothesis $H_0 : \lambda = \lambda_0$ for some $\lambda_0 > 0$. Against alternatives $H_1 : \lambda < \lambda_0$ or $H_1 : \lambda > \lambda_0$, the likelihood ratio test statistic is*

$$(\lambda_0/\lambda)^n \exp\left\{-(\lambda_0 - \lambda) \sum_{i=1}^n t_i\right\}. \tag{5}$$

*By monotonicity, testing (5) is equivalent to testing upper or lower tail probabilities of $\sum_{i=1}^n t_i$. More specifically, under $H_1 : \lambda < \lambda_0$, corresponding to an alternative of stochastically larger inter-arrival times, low values of (5) correspond to high values $\sum_{i=1}^n t_i$. Under this $H_1$, the individual p-value for the $i$th observation would be the survivor function for $t_i$ under $H_0$,*

$$p_i = \exp(-\lambda_0 t_i). \tag{6}$$

*Rearranging (6), $t_i = -\log(p_i)/\lambda_0$ so testing the magnitude of $\sum_{i=1}^n t_i$ is equivalent to testing $\sum_{i=1}^n \log p_i$, which is Fisher's method. Finally, under $H_1$ the observation $t_i$ has density $\bar{f}_1(t_i) = \mathbb{I}_{[0,\infty]}(t_i) \lambda \exp(-\lambda t_i)$ and so by the change of variable (6), $f_1(p_i) \propto \mathbb{I}_{[0,1]}(p_i) p_i^{\lambda/\lambda_0 - 1}$ and hence $p_i \sim \text{Beta}(\lambda/\lambda_0, 1)$ where $\lambda/\lambda_0 < 1$.*

*Conversely, if $H_1 : \lambda > \lambda_0$, implying stochastically smaller inter-arrival times, then the relevant p-value is $p_i = 1 - \exp(-\lambda_0 t_i)$. Testing the lower tail of the likelihood ratio*

(5) is then equivalent to calculating the lower tail probability of $\sum_{i=1}^{n} \log(1 - p_i)$, which corresponds to Pearson's method, and under $H_1$, $p_i \sim \text{Beta}(1, \lambda/\lambda_0)$ where $\lambda/\lambda_0 > 1$.

**Proposition 2.** *Consider an alternative hypothesis for p-values,*

$$H_1 : p_{i^*} \sim \text{Beta}(a, n), \text{ for some } i^* \in \{1, \ldots, n\}, \quad p_i \sim \text{U}[p_{i^*}, 1], \quad i = 1, \ldots, n, \ i \neq i^*$$

*with $a \in (0, 1)$. Then the optimal p-value combination method is $S_T$.*

*Proof.* As $p_{i^*}$ must be the minimum $p$-value, the likelihood ratio yields the result. □

Under $H_0$, $p_{i^*}$ would have a $\text{Beta}(1, n)$ distribution, so $a < 1$ implies a surprisingly small minimum $p$-value. However, the remaining details of $H_1$ are uncomfortable, as $\text{U}(p_{i^*}, 1)$ $p$-values imply a departure from the assumption of independence with an unusual dependency structure: the test statistics are draws from their null models but truncated to be less extreme than the $i^*$th test statistic. For this reason, Tippett's method seems very difficult to justify in practice.

## 2.2 Truncated Gamma Distribution

Another natural choice for an alternative $p$-value density is to truncate to $[0, 1]$ a continuous distribution on the positive half-line. Denoting by $\Gamma_{[0,1]}(a, b)$ the gamma distribution truncated to the unit interval with density $f_1(p) \propto \mathbb{I}_{[0,1]}(p) \, p^{a-1} \exp(-bp)$, $a, b > 0$; then $f_1(p)$ is non-increasing provided $a \in (0, 1]$.

**Proposition 3.** *Consider an alternative hypothesis for p-values, $H_1 : p_i \sim \Gamma_{[0,1]}(a, b)$, $i = 1, \ldots, n$, with $a \in (0, 1]$. Let $w = (1 - a)/(1 + b - a)$. Then the uniformly most powerful test statistic for combining $p_1, \ldots, p_n$ is of the form*

$$w \sum_{i=1}^{n} \log p_i + (1 - w) \sum_{i=1}^{n} p_i. \tag{7}$$

The proof is similar to Proposition 1. Note the similarity of (4) and (7) for small $p$-values. Moreover, if $a = 1$ so the $p$-values have a truncated exponential distribution $\text{Exp}_{[0,1]}(b) \equiv \Gamma_{[0,1]}(1, b)$ under $H_1$, then $w = 0$ and Edgington's method of summing $p$-values is optimal.

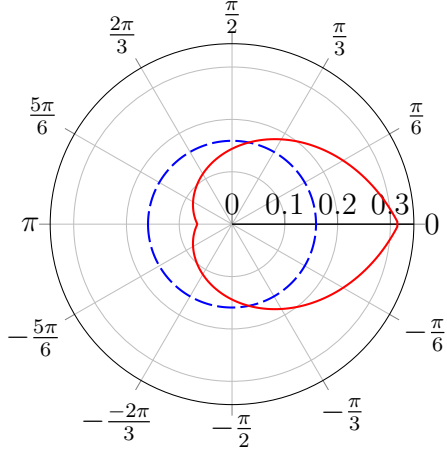

Figure 3: Uniform$[-\pi, \pi)$ (- -) and Wrapped Double Exponential(0.5) (——) densities.

**Example 2.2.** *Suppose $t_1, \ldots, t_n$ are the angles in $[-\pi, \pi)$ of $n$ independently distributed random points lying on the unit circle. These could be the event times of a counting process modulo some fixed period, such as the time of day of event times. Further suppose $H_0: t_i \sim \mathrm{U}[-\pi, \pi)$, $H_1: t_i \sim$ Wrapped Double Exponential($\lambda$) where under $H_1$, $t_i$ has density*

$$\bar{f}_1(t_i) = \mathbb{I}_{[-\pi,\pi)}(t_i) \lambda \exp(-\lambda |t_i|) / [2\{1 - \exp(-\lambda \pi)\}], \ \lambda > 0.$$

*Then $H_0$ represents the maximum entropy distribution on the circle, whilst $H_1$ represents the maximum entropy distribution on the circle having zero mean; the two densities are depicted in Figure 3. These hypotheses could arise if comparing a homogeneous Poisson process model against an inhomogeneous Poisson process model with a periodic intensity function which has a symmetric exponential rise and fall within each period.*

*The likelihood ratio test statistic simplifies to*

$$[\{1 - \exp(-\lambda \pi)\}/\lambda \pi]^n \exp\left(\lambda \sum_{i=1}^n |t_i|\right),$$

*which is an increasing function of $\sum_{i=1}^n |t_i|$. Since the density $\bar{f}_1$ is decreasing in either direction from zero, the individual p-value for $t_i$ is the probability under $H_0$ of lying even closer to zero,*

$$p_i = \int_{t:|t| \leq |t_i|} \frac{\mathrm{d}t}{2\pi} = \frac{|t_i|}{\pi}.$$

*Hence a test of the sum of the p-values, Edgington's method, is the likelihood ratio test.*

## 2.3 General Distribution Functions

For a general approach to devising alternative hypothesis densities for $p$-values, consider the following hypothesis test for independent test statistics $t_1, \ldots, t_n$: $H_0 : t_i \sim \bar{F}_0$, $H_1 : t_i \sim \bar{F}_1$, where $\bar{F}_0$ and $\bar{F}_1$ are any two absolutely continuous distribution functions with common support. Suppose draws from $\bar{F}_1$ are stochastically smaller than draws from $\bar{F}_0$, implying an individual test can be derived from the lower-tail $p$-value for the observation $t_i$, $p_i = \bar{F}_0(t_i)$.

Under $H_0$, clearly $p_i \sim \mathrm{U}[0,1]$. Under $H_1$,

$$\mathrm{pr}_{H_1}(p_i \leq p) = \mathrm{pr}_{H_1}\{\bar{F}_0(t_i) \leq p\} = \mathrm{pr}_{H_1}\{t_i \leq \bar{F}_0^{-1}(p)\} = \bar{F}_1\{\bar{F}_0^{-1}(p)\}$$
$$f_1(p) = \bar{f}_1\{\bar{F}_0^{-1}(p)\}/\bar{f}_0\{\bar{F}_0^{-1}(p)\},$$

where $\bar{f}_0$ and $\bar{f}_1$ are the respective densities of $\bar{F}_0$ and $\bar{F}_1$. Thus the density of the $p$-values under $H_1$ is non-increasing, and therefore admissible, if and only if the ratio of densities $\bar{f}_1/\bar{f}_0$ is non-increasing.

**Example 2.3.** *Suppose $t_1, \ldots, t_n$ are independent $\mathrm{N}(\mu, 1)$ variables. Consider the one-sided test: $H_0 : \mu = 0$, $H_1 : \mu < 0$. Clearly the ratio $\bar{f}_1(t)/\bar{f}_0(t) \propto \exp(\mu t)$ is non-increasing under $H_1$. The likelihood ratio test in this setting is the familiar one-sided test of a Gaussian mean, with test statistic $\sum_{i=1}^{n} t_i$. And here, $t_i = \Phi^{-1}(p_i)$, implying that Stouffer's method is most powerful.*

*The density of the $p$-values under $H_1$ is $f_1(p) = \bar{f}_1\{\Phi^{-1}(p)\}/\bar{f}_0\{\Phi^{-1}(p)\} \propto \exp\{\mu\Phi^{-1}(p)\}$. The right panel of Figure 2 shows this density for $\mu = -1$; note the similarity with the beta distribution which had been proved to be suited to George's method, which was noted in Section 1 to well approximate Stouffer's method.*

## 2.4 Weighted Meta-Analyses

In some circumstances it can be desirable to attribute different weights to the $p$-values being combined in a meta-analysis. For example, if the sample sizes of the underlying studies varied considerably, weights $w_1, \ldots, w_n$ might be chosen proportional to the respective sample sizes, or their square roots.

Stouffer's method is commonly preferred in the presence of weights, as the weighted test statistic $\sum_{i=1}^{n} w_i \Phi^{-1}(p_i)$ retains a closed form distribution, $\mathrm{N}(0, \sum_{i=1}^{n} w_i)$, under the null hypothesis (1). Following on from Example 2.3, this statistic is the most powerful combiner of $p$-values derived from tests of $H_0 : t_i \sim \mathrm{N}(0, w_i^{-2})$, $H_1 : t_i \sim \mathrm{N}(\mu, w_i^{-2})$, $\mu \neq 0$.

Through similar arguments from earlier sections, the following results are easily established for the other $p$-value combiners: A weighted Fisher's method $\sum_{i=1}^n w_i \log p_i$ is optimal for $p$-values under the rival hypotheses $H_0 : p_i \sim \mathrm{U}(0,1)$, $H_1 : p_i \sim \mathrm{Beta}(w_i a, 1)$, $0 < a < 1$. A weighted Pearson's method $-\sum_{i=1}^n w_i \log(1 - p_i)$ is optimal under $H_0 : p_i \sim \mathrm{U}(0,1)$, $H_1 : p_i \sim \mathrm{Beta}(1, w_i b)$, $b > 1$. A weighted Edgington's method $\sum_{i=1}^n w_i p_i$ is optimal under $H_0 : p_i \sim \mathrm{U}(0,1)$, $H_1 : p_i \sim \mathrm{Exp}_{[0,1]}(w_i b)$, $b > 1$. It follows from Example 2.1 that the weighted versions of Fisher's and Pearson's methods might naturally arise as the optimal combiners of $p$-values from inter-arrival time or event time data when the alternative hypothesis assumes a time-varying hazard rate or intensity.

Although the null distributions of these other weighted $p$-value combiners do not have a convenient form, their first two moments are trivial to calculate and the central limit theorem implies asymptotic Gaussianity in all cases, and so practical implementation remains straightforward. For any particular combination method, the Berry—Esseen theorem (Berry, 1941) suggests the rate of convergence to normality depends on the so-called effective sample size implied by the weights,

$$\left(\sum_{i=1}^n w_i\right)^2 \Big/ \sum_{i=1}^n w_i^2. \tag{8}$$

For the weighted version of George's method, for example, the distribution of $\sum_{i=1}^n w_i \log\{p_i/(1-p_i)\}$ can be shown (Hedges and Olkin, 1985) to have an approximate Student's $t$-distribution with the degrees of freedom parameter an increasing function of (8).

The theory above gives the optimal methods for combining $p$-values in some particularly tractable cases. By the continuity of the $p$-value combination methods considered, it follows that the same methods will be near-optimal for distributions which are similar to these tractable cases. On this basis, Table 1 proposes an informal rule-of-thumb for choosing a $p$-value combination method, based on the underlying data types and tests that gave rise to the $p$-values.

Table 1: Choices of $p$-value combination methods from the underlying data types and tests.

| Data/test type | Method |
|---|---|
| Positive-valued data, larger under $H_1$ | Fisher |
| Positive-valued data, smaller under $H_1$ | Pearson |
| Real-valued, approximately Gaussian data | George/Stouffer |
| Circular data | Edgington |

## 3 Example: Meta-analysis of F-tests

The F-distribution and associated F-tests are routinely used in various model selection contexts, such as tests of equal variances, analysis of variance, and regression analysis. The F-distribution is not an exponential family and the $p$-values from F-tests derive from regularised incomplete beta functions, so performing a meta analysis combining F-tests is not straightforward; see Prendergast and Staudte (2016). Therefore $p$-values derived from F-distributions are an interesting test case for examining the performance of different $p$-value combiners. Let $F_{\nu_1,\nu_2}$ denote the F-distribution with $\nu_1$ and $\nu_2$ degrees of freedom, and here suppose $\nu_1 = 1$ and $t_1, \ldots, t_n$ are $n > 1$ independent draws from $F_{1,\nu_2}$.

First suppose rival hypotheses $H_0 : \nu_2 = 2$, $H_1 : \nu_2 = 1$. Draws from the alternative $F_{1,1}$ are stochastically larger than those from $F_{1,2}$, and so the $F_{1,2}$ upper-tail probability of each $t_i$ gives a relevant null distribution $p$-value. The top row of Figure 4 shows the power curves when combining these $p$-values from $n$ tests for which $H_1$ is true, for $n = 2, 10, 50, 100$, derived from 100,000 simulations. Fisher's method is most appropriate in this setting, and Pearson's method considerably inferior. Example 2.1 and Table 1 showed that Fisher's method was optimal for combining $p$-values for exponentially distributed waiting times which were stochastically larger than anticipated under the null. In the present setting, the F-distribution yields positive-valued test-statistics, and Fisher's test shows the best performance when these statistics are surprisingly large.

Similarly, if the null and alternative hypotheses are reversed to $H_0 : \nu_2 = 1$, $H_1 : \nu_2 = 2$, such that the positive-valued $t_i$ are stochastically smaller under the alternative hypothesis, then Table 1 suggests that Pearson's method should be near-optimal. The bottom row of

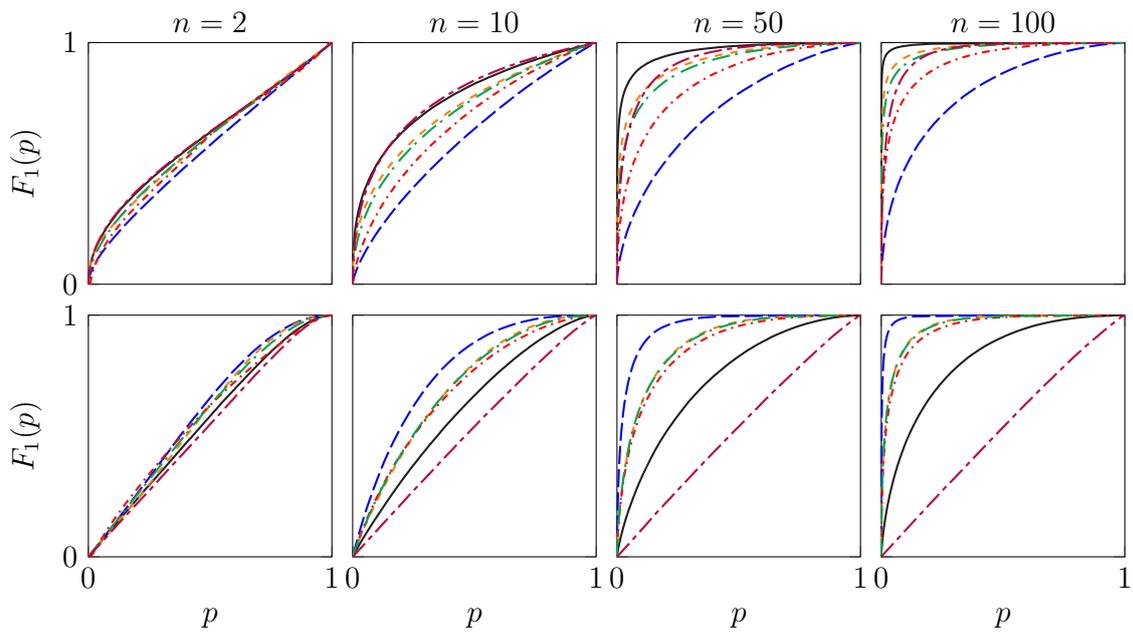

Figure 4: Distribution of meta-analysis $p$-values arising from combining $n$ F-tests with larger (top row) or smaller (bottom row) than expected test statistics, using Fisher's method (——), Pearson's method (– –), George's method (- - -), Edgington's method (-·-·), Stouffer's method (—·-) and Tippett's method (—·-).

Figure 4 shows the corresponding power curves, with Pearson's method the clear winner and Fisher's method now performing poorly. Tippett's method is effectively unable to distinguish $H_1$ from $H_0$.

The similarity of the power curves of Edgington's method across the two cases suggests that the summing of $p$-values provides a reasonably robust combiner if $p$-values are from a mixture of these two alternatives. In addition, the power curves from George's and Stouffer's methods are above those from Edgington's method, and are either very close together or indistinguishable, confirming remarks in Section 1 about the known similarity of these two methods. The improved robustness of George's method against Edgington's might be expected, with George's method a hybrid of the two optimal combiners under the two alternatives; Edgington's method, on the other hand, as seen in Example 2.2 is well-suited to $p$-values derived from circular data.

# Acknowledgement

The research for this article was funded by the Heilbronn Institute for Mathematical Research.